\documentclass[aoas,MSNbibl,nameyear,dvips]{arximspdf}
\usepackage{multirow,dcolumn}
\usepackage{graphicx}

\doi{10.1214/13-AOAS666} 
\volume{7}
\issue{4}
\pubyear{2013}
\firstpage{2249}
\lastpage{2271}

\makeatletter
\newproclaim{algo}{Algorithm}
\newcolumntype{d}[1]{D{.}{.}{#1}}
\newcommand{\eqref}[1]{(\ref{#1})}
\newtheorem{theorem}{Theorem}
\newtheorem{lemma}{Lemma}

\newcommand{\bI}{\mathcal{I}}
\newcommand{\bt}{\mathbf{t}}
\newcommand{\bx}{\mathbf{x}}
\makeatother

\begin{document}
\begin{frontmatter}

\title{How do heterogeneities in operating environments affect field
failure predictions and test planning?}

\begin{aug}
\author[A]{\fnms{Zhi-Sheng} \snm{Ye}\corref{}\thanksref{m1,t1}\ead[label=e1]{iseyez@gmail.com}},
\author[B]{\fnms{Yili} \snm{Hong}\thanksref{m2,t2}}
\and
\author[B]{\fnms{Yimeng} \snm{Xie}\thanksref{m2,t2}}
\runauthor{Z.-S. Ye, Y. Hong and Y. Xie}
\affiliation{Hong Kong Polytechnic University\thanksmark{m1} and
Virginia Tech\thanksmark{m2}}
\thankstext{t1}{Supported by the Sate Key Laboratory of Industrial
Control Technology with project code ICT1313.}
\runtitle{Heterogeneous environments and field failure predictions}
\thankstext{t2}{Supported by NSF Grant CMMI-1068933 to
Virginia Tech and the 2011 DuPont Young Professor Grant.}
\address[A]{Z.-S. Ye\\
Department of Applied Mathematics\\
Hong Kong Polytechnic University\\
Kowloon\\
Hong Kong\\
\printead{e1}} 
\address[B]{Y. Hong\\
Y. Xie\\
Department of Statistics\\
Virginia Tech\\
Blacksburg, Virginia 24061\\
USA}
\end{aug}

\received{\smonth{12} \syear{2012}}
\revised{\smonth{6} \syear{2013}}

%
\begin{abstract}
The main objective of accelerated life tests (ALTs) is to predict
fraction failings of products in the field.
However, there are often discrepancies between the predicted fraction
failing from the lab testing data and that from the field failure data,
due to the yet unobserved heterogeneities in usage and operating
conditions. Most previous research on ALT planning and data analysis
ignores the discrepancies, resulting in inferior test plans and biased
predictions. In this paper we model the heterogeneous environments
together with their effects on the product failures as a frailty term
to link the lab failure time distribution and field failure time
distribution of a product. We show that in the presence of the
heterogeneous operating conditions, the hazard rate function of the
field failure time distribution exhibits a range of shapes. Statistical
inference procedure for the frailty models is developed when both the
ALT data and the field failure data are available. Based on the frailty
models, optimal ALT plans aimed at predicting the field failure time
distribution are obtained. The developed methods are demonstrated
through a real life example.
\end{abstract}

%
\begin{keyword}
\kwd{Accelerated life test data}
\kwd{frailty model}
\kwd{field failure data}
\kwd{heterogeneous operating conditions}
\kwd{optimal plan}
\end{keyword}

\end{frontmatter}

\section{Introduction}\label{secintro}
\subsection{Motivation}\label{secmotivation}
Most commercial products are sold with warranties. Before a new product
is launched to the market, it is extremely important to accurately
estimate the proportion of field returns within a given warranty period
in order to determine the monetary reserves for covering future
warranty claims. The failure information can be obtained through
pre-launch accelerated life tests (ALTs) in a timely fashion. In an
ALT, a~number of samples\vadjust{\goodbreak} are tested under harsh conditions, for
example, a combination of high voltage, temperature, pressure, use
rate, etc., which yields information on product reliability within a
reasonable time frame. Failure time data from the test are collected,
analyzed and extrapolated to estimate lifetime characteristics of
interest at \textit{nominal} use conditions based on some stress-life
models. There is a bulk of literature on ALT data analysis and optimal
design of ALT experiments. See \citet{pascual06}, \citet{ma08},
\citet
{guo11} and \citet{liu12}, among others. The use conditions are
implicitly assumed to be homogeneous (same for all customers) in most
ALT research, including the above references.

After the product is sold to customers with a warranty, units that fail
within the warranty period are returned to the manufacturer for repair
or replacement, which are known as warranty claims. These warranty
claim data reflect failure behaviors of the product under \textit{actual}
use conditions. Analysis of these warranty return data is useful
because it validates the results from ALT data analysis, and can be
used to improve the accuracy of parameter estimation from the ALT. See
\citet{blischke11} for an overview of this topic.

However, large discrepancies between the results of ALT data analysis
and field failure data analysis are often found. Analysis of field
failure data tends to suggest higher variability in the product's
failure times compared with the result based on ALT data analysis.
Conceivably, this is because products in the field are usually exposed
to heterogeneous usage and operating conditions. A motivating example
is as follows.

\citet{567} described an application involving an appliance, which is
called Appliance B. Appliance B contains a turbine device which has two
major failure modes: crack failure modes and wear failure modes.
Engineering knowledge suggests that it is reasonable to assume that
these two failure modes are independent. For illustration, we only
consider the wear failure mode, accounting for around 80\% of the total
field failures. Appliance B was sold with a two-year warranty. Before
its entry into the market, an ALT was conducted to obtain reliability
information of the product, in which 10 units were subject to a wear
test. Field failure data were also available during the subsequent
warranty tracking study of 4708 units with 93 wear failures. More
details can be found in \citet{567}.

According to the analysis in Section~\ref{secexample}, the Weibull
distribution provides a good fit to the failure data from ALT, but it
does not
provide an adequate fit to the field data. As we will argue, the
discrepancy is largely due to the varying operating conditions in the
field. When varying operating conditions are taken into account, theory
suggests the use of other distributions for the field data, such as the
Burr-XII distribution, which do fit well.

\subsection{Heterogeneous operating conditions}
The operating conditions are dynamic in a number of ways.\vadjust{\goodbreak} First off,
products are used in different geographical areas because of
customer
locations. Therefore, the operating environments (e.g., temperature,
humility, etc.) are heterogeneous for units across the product
population. Second, different users have different usage behaviors. In
a two-dimensional warranty analysis, it is commonly assumed that the
use rate of a customer is constant and it varies across the customer
population [\citet{485,884}]. \citet{752} also observed that the field
stress level may vary over the product population. Moreover, the usage
profile can be time dependent. As an example, \citet{830} reported a
problem where the stress profile, for example, pressure and
temperature, over time for a seal in brake cylinders is stochastic. The
presence of variable operating conditions significantly influences
failures of the product. As suggested from consumer reports in February
1991 [\citet{784}], the percentage of washer--dryer machines that ended
up with a warranty claim went up from 14\% among those who reported an
average of one to four laundry loads per week to 25\% among those who
reported an average in excess of eight loads per week. Furthermore,
this pattern was observed across brands consistently.

In the presence of heterogeneous operating conditions of the product
population, direct prediction of the proportion of warranty returns
from ALT data analysis can be highly biased. In principle, the failure
time distribution of the in-lab testing units can be linked to that of
the field population by taking into account information about these
dynamics in environments. The information includes the types of
significant dynamic environmental factors, the distributions for these
factors as well as the acceleration relationships that relate each
factor to the failure process. Among these environmental factors,
information about the use rate may be the easiest to collect. For
example, \citet{567} and \citet{752} focused on modeling the effects of
usage rates. Both studies assumed a constant usage rate for an
individual unit and a lognormal distribution for usage rates across the
product population. However, the field failure time distribution in
\citet{752} does not have closed-form expressions, which makes analysis
of field return data and verification of model assumptions (e.g., the
lognormal assumption of the usage rate distribution) very difficult,
and which greatly complicates the ALT planning for a new vintage of the
product under similar environments. Even if the distribution of the
usage rate is available, say, from a customer survey, the models in
these two studies still ignore other influential factors such as
heterogeneous customer locations. In fact, it is almost impossible to
directly collect information (i.e., distributions for each
environmental factor and their respective effects on the failure
process) about all heterogeneous environmental factors other than the
usage rate.

\subsection{Objectives and overview}\label{secobjective}
This paper is an endeavor to answer the question of how heterogeneities
in operating environments affect predictions of field failures and
planning\vadjust{\goodbreak} of ALTs. We treat the \textit{unobservable operating factors} as
well as \textit{their effects} on the product failure process as a
``frailty,'' through which the lab failure time distribution of a
product can be linked to the field failure time distribution. The
``frailty'' is an unobservable random variable used to account for
heterogeneities caused by unobservable covariates. In its simplest
form, the frailty is an unobserved random proportionality factor that modifies
the baseline failure rate function of an individual, which is similar
to the multiplicative effect of a covariate on the failure rate in
Cox's proportional hazard model. In biostatistics, lifetime models with
frailties have attracted much attention, for example, see \citet
{Hanagal11} for a book length treatment on this area. In reliability
engineering, the frailty is often called a random effect and also
receives some applications, for example, see \citet{stefanescu06},
\citet{697} and \citet{939}, among others. However, one challenge of
using frailty is that the resulting marginal distribution is often
mathematically intractable.

This paper develops tractable frailty models that relate ALT failures
to warranty failures. We show that in the presence of the frailty, the
hazard rate of a field unit exhibits various shapes. An appropriate
distribution for the frailty can be determined through joint modeling
of both ALT data and warranty return data. Detailed procedures to
analyze the data and to collate the frailty distribution are developed.
The results enable the prediction of field failures for a future
product through analysis of ALT data. We also derive optimal designs of
ALT experiments for a new vintage and show how the heterogeneities
affect the optimal ALT design.

The remainder of the paper is organized as follows. Section~\ref{secmodel} introduces the gamma frailty model for linking lab test
data and field failure data and investigates possible shapes of the
field failure rate. In Section~\ref{secinference} a procedure for
statistical inference of the frailty model is developed. We also
extensively discuss the model validation through hypothesis testing.
Optimal ALT plans under the frailty model are obtained in Section~\ref{secoptimalALT}. Section~\ref{secexample} applies the frailty model
to the Appliance B example. Section~\ref{secconclusion} concludes the paper.

\section{Linking lab failures and field failures}\label{secmodel}
Under the stable lab testing conditions, we assume the lifetime $X$ of
the product follows a Weibull distribution, which is one of the most
commonly used lifetime distributions. However, existence of the
heterogeneous operating conditions influences lifetime of a field unit.
The basic idea is to introduce into the hazard rate an additional
random parameter $Z$ that accounts for the heterogeneities. The frailty
$Z$ links the distribution of $X$ to that of the field failure time
$T$. In this section, the frailty model is developed and the hazard
rate of $T$ is investigated.

\subsection{Failures in lab testing}\label{seclabcdf}
As suggested by the extreme value theory, the Weibull distribution is
an appropriate lifetime model when the failure is caused by the
weakest\vadjust{\goodbreak}
flaw/link in a unit. It has been widely used for modeling lifetime of
products and components. The failure time $X$ of a lab testing unit is
assumed to follow a Weibull distribution with the respective cumulative
distribution function~(c.d.f.) and probability density function (p.d.f.)
given by
\[
F_X(x) = 1 - \exp \biggl[- \biggl(\frac{x}{\alpha}
\biggr)^\beta \biggr],\qquad x>0
\]
and
%
%
\begin{equation}
\label{eqnweibull} f_X(x) = \frac{\beta}{\alpha} \biggl(
\frac{x}{\alpha} \biggr)^{\beta-1} \exp \biggl[- \biggl(\frac{x}{\alpha}
\biggr)^\beta \biggr],\qquad x>0,
\end{equation}
where $\alpha>0$ is the scale parameter and $\beta>0$ is the shape
parameter. The hazard rate function of $X$ is given by
%
%
\begin{equation}
h_X(x) = \frac{\beta}{\alpha} \biggl(\frac{x}{\alpha}
\biggr)^{\beta-1}.
\end{equation}
It is well known that the hazard rate function is monotone increasing
when $\beta>1$ and monotone decreasing when $0<\beta<1$.

\subsection{Field failures: A gamma frailty model}
When the product is sold to customers, the operating conditions are
heterogeneous and unobservable. The unobservable effects are described
by a frailty $Z$. The frailty $Z$ is constant for a unit and varies
across the product population. Conditional on $Z$, the lifetime of a
field unit follows the Weibull distribution with a hazard rate function
given by
%
%
\begin{equation}
\label{eqnph} h_T(t; Z)=Zh_X(t)= Z\times
\frac{\beta}{\alpha} \biggl(\frac
{t}{\alpha
} \biggr)^{\beta-1}.
\end{equation}
Because the baseline distribution is Weibull, this frailty model is
similar to assuming a random scale parameter $\alpha$ [\citet{416}, page
457]. Previously, \citet{567} and \citet{752} adopted such method to
accommodate information on the heterogeneities. However, the reason we
do not use a random scale parameter is that it is difficult, if not
impossible, to find a distribution for $\alpha$ such that the resulting
field failure time distribution has a closed form.

The distribution of $Z$ depends on the heterogeneities of the field
environments as well as the effects of the random environments on the
product. For example, when the heterogeneities are caused by the random
use rate $U$, previous research suggests that the effect of $U$ on
product failures can be empirically described by a power law relation
$Z=aU^b$, $a, b >0$ are parameters, while the use rate distribution
tends to be unimodal and positively skewed. This leads to a unimodal
and positively skewed distribution for $aU^b$. Therefore, distributions
like the gamma [\citet{264,485}], lognormal [\citet{475,567}] and
inverse Gaussian distributions are appropriate for $Z$. Occasionally,
the uniform distribution is also recommended [\citet{266}]. The frailty
$Z$ includes the random usage rate and, thus, it is reasonable to
assume that it is also unimodal and positively skewed. To specify a
distribution family for the frailty $Z$, it is of advantage that the
resulting field failure distribution is tractable. This is because when
the distribution of $Z$ has a closed form, we can easily collate the
validity of the frailty distribution through data analysis. We find
that the families of gamma, inverse Gaussian and uniform distributions
for the frailty result in tractable distributions for $T$. In the
motivating example described in Section~\ref{secmotivation}, the
frailty is found to be well described by the gamma distribution.
Therefore, this paper focuses on the gamma frailty model. \emph
{Development of the inverse Gaussian frailty model and the uniform
frailty model is put in the supplemental material} [\citet
{ye13supplement}]. In fact, as suggested by \citet{627}, the gamma
distribution is highly flexible to reflect p.d.f.s of most shapes and,
thus, the gamma frailty model is applicable to similar problems other
than the Appliance B example.

In this section, we consider the gamma distribution with a threshold
parameter in order to demonstrate the fact that the hazard rate
function of $T$ exhibits various shapes. The three-parameter gamma
distribution with a threshold parameter $\gamma$ has a p.d.f. given by
%
%
\begin{equation}
\label{eqn3Pgamma} \varphi(z) = \frac{\mu^k(z-\gamma)^{k-1}}{\Gamma(k)} \exp \bigl[-\mu (z-\gamma)
\bigr], \qquad z>\gamma.
\end{equation}
When the frailty follows a distribution specified by (\ref
{eqn3Pgamma}), it can be shown by marginalizing $Z$ out of (\ref
{eqnph}) that the c.d.f. and p.d.f. of $T$ are, respectively, given by
%
%
\begin{eqnarray}
\label{eqnfieldDistrib} F_T(t) &=& 1 - \bigl[(t/
\alpha)^\beta/\mu+ 1 \bigr]^{-k} \exp \bigl[-\gamma(t/
\alpha)^\beta \bigr],
\nonumber
\\
f_T(t) &=& \frac{\beta}{\alpha} \biggl(\frac{t}{\alpha}
\biggr)^{\beta-1} \biggl[\frac{(t/\alpha)^\beta}{\mu} + 1 \biggr]^{-k} \biggl\{
\gamma +k \biggl[ \biggl(\frac{t}{\alpha} \biggr)^\beta+ \mu
\biggr]^{-1} \biggr\} \\
&&{}\times\exp \biggl[-\gamma \biggl(\frac{t}{\alpha}
\biggr)^\beta \biggr].\nonumber
\end{eqnarray}

It is interesting to note that when $\gamma= 0$, model (\ref
{eqnfieldDistrib}) reduces to the Burr-XII distribution. The Burr-XII
distribution has been used in reliability analysis by a few
researchers, for example, see \citet{700,953,952} and \citet{951}, to
name a few. However, the Burr-XII distribution is much less popular
than the lognormal distribution. Nevertheless, this distribution has
several advantages over the lognormal distribution. Similar to the
lognormal distribution, the Burr-XII distribution also has a unimodal
hazard rate. But compared with the lognormal distribution, the Burr-XII
distribution is more flexible in analysis of survival data. For
example, parameters of the Burr-XII distribution can be determined
through a simple probability plotting procedure [\citet{700}]. In
addition, it has greater mathematical tractability when dealing with
censored data which are very common in lifetime data analysis. The
contribution of a right-censored observation to the likelihood is equal
to the value of the survival function at the time of censoring, which
can be evaluated explicitly for the Burr-XII distribution, but not for
the log-normal distribution.

When $\gamma= 0$, the mean and variance of the frailty variable $Z$
are $k/\mu$ and $k/\mu^2$, respectively. If we fix $k/\mu$ at a
constant and let $\mu\rightarrow\infty$, then the distribution of $Z$
will degenerate to a single point, and the Burr-XII distribution will
also degenerate to a Weibull distribution. This is legitimate because
under such circumstance, there is no variation in the frailty. The
log-logistic distribution, a common distribution used in lifetime data
analysis, is also a special case of model (\ref{eqnfieldDistrib}),
when $\gamma= 0$ and $k=1$.

\subsection{Hazard rate for units in the field}
In reliability assessment, the shape of the hazard rate reflects the
early failure and aging behavior of the product. Therefore, it is
important to know the shape with a view to scheduling preventive
maintenance and detecting possible early failure modes. The hazard rate
function of~$Z$ can be readily obtained by dividing the p.d.f. by the
survival function, that is, $1 - F_T(t)$, which gives
%
%
\begin{equation}
\label{eqnfieldFailureRate} h_T(t) = \frac{\gamma\beta}{\alpha} \biggl(
\frac{t}{\alpha} \biggr)^{\beta
-1} + \frac{k\beta t^{\beta-1}}{t^\beta+ \mu\alpha^\beta}.
\end{equation}
The hazard rate of this distribution exhibits various shapes, as can be
checked through the first order derivative of (\ref
{eqnfieldFailureRate}) with respect to $t$. By and large, the hazard
rate could have four possible shapes, as summarized below.

\begin{longlist}[\textit{Case} 1.]
\item[\textit{Case} 1.] $\beta\leq1$.

The hazard rate $h_T(t)$ is decreasing in $t$. Specifically, when
$\beta< 1$, $h_T(t)$ decreases from $\infty$ to 0. When $\beta= 1$,
$h_T(t)$ decreases from $\gamma+ k/\mu$ to $\gamma$. This is because a
mixture of distributions with decreasing hazard rates has a
nonincreasing hazard rate.

\begin{figure}

\includegraphics{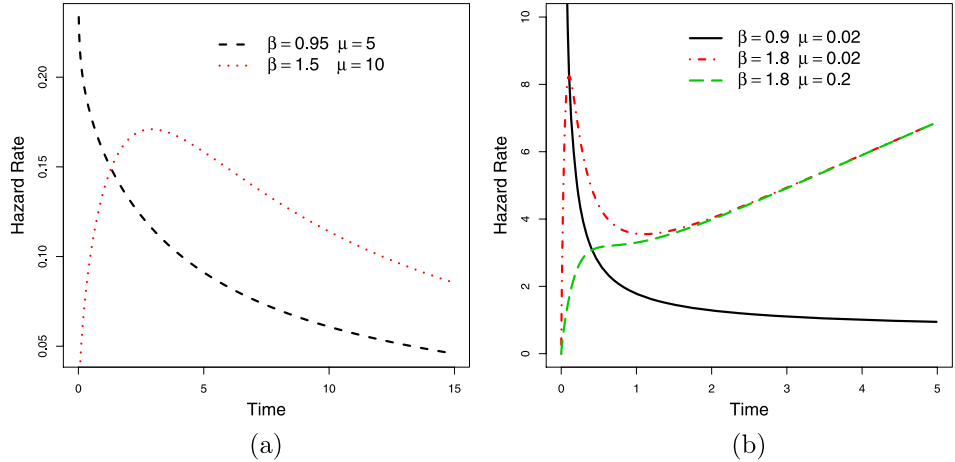}

\caption{Illustrations of some shapes of the hazard function in
\protect\eqref{eqnfieldFailureRate}:
\textup{(a)} $\gamma=0, \alpha=1, k=1$; and \textup{(b)}~$\gamma=1, \alpha=1, k=1$.}
\label{figfailurerate}
\end{figure}

\item[\textit{Case} 2.] $\gamma> 0, \beta>1, \beta^2-\beta<\frac
{k}{4\gamma\mu}$.

The hazard rate $h_T(t)$ exhibits an N-shape.

\item[\textit{Case} 3.] $\gamma> 0, \beta>1, \beta^2-\beta>\frac
{k}{4\gamma\mu}$.

The hazard rate $h_T(t)$ is increasing.

\item[\textit{Case} 4.] $\gamma= 0$ and $\beta> 1$.

The hazard rate $h_T(t)$ has an upside-down bathtub shape.
\end{longlist}

Some typical curves of the hazard rate are depicted in Figure~\ref{figfailurerate}. It is interesting to see that when $\beta>1$, the
hazard rate under lab conditions is increasing, but the hazard rate of
a field unit can be either increasing, unimodal or N-shape. When the
hazard rate of $T$ is unimodal or N-shape, the initial hazard rate can
be very high, as can be seen from the dash dotted lines in Figure~\ref{figfailurerate}. In practice, when a manufacturer observes a high
hazard rate at the early stage, he may suspect that it is the infant
mortality caused by defects. The analysis in this section reveals that
early failures can also be caused by units operated under harsh
environments (i.e., large realizations of $Z$). These units are more
likely to fail and, hence, more ``frail'' than other field units.

%

\section{Statistical inference}\label{secinference}
Information about the distribution of the frailty can be obtained
through a joint analysis of lab data and field data. In the previous
section we adopt the three-parameter gamma distribution with a
threshold parameter~$\gamma$ for the frailty $Z$ to set forth the fact
that the field hazard rate can have various shapes in the presence of
heterogeneous operating conditions. In reality, the frailty $Z$ often
ranges from zero to infinity. Thus, this section focuses on the case
when the frailty follows a regular two-parameter gamma distribution
(i.e., $\gamma= 0$), under which the field failure time $T$ follows
the Burr-XII distribution.

Suppose that $n$ units are tested in the lab and $x_i$ is the observed
failure time or censoring time for the $i$th unit. Further, let $\delta
_i$ be the censoring indicator, where $\delta_i=0$ when the unit is
right censored and 1 otherwise. Therefore, for the $i$th lab unit, we
observe $(x_i,\delta_i)$. Similarly, suppose we observe the
failure\vadjust{\goodbreak}
times of $N$ field units $(t_j, \tilde\delta_j)$, $j=1,2,\ldots,N$,
where the field-data censoring indicator $\tilde\delta_j=0$ when the
$j$th unit is right censored and 1 if it fails and is returned as a
warranty claim.

\subsection{Estimation and hypothesis tests}\label{secestimation}
Given the lab testing data and the warranty return data for the same
product, we develop a procedure to analyze the data by capitalizing on
the model in Section~\ref{secmodel}. In this procedure, we need to
first collate the Weibull distribution (\ref{eqnweibull}) for the ALT
data, and then check if the warranty return data conform to the
Burr-XII distribution with c.d.f.
%
%
\begin{equation}
\label{eqnburrXII} G(t) = 1 - \bigl[(t/\lambda)^{\beta} + 1
\bigr]^{-k}, \qquad t > 0.
\end{equation}
It is noted from (\ref{eqnweibull}) and (\ref{eqnburrXII}) that when
the gamma frailty model holds, the Weibull shape parameter in (\ref
{eqnweibull}) should be equal to $\beta$ in the Burr-XII distribution
(\ref{eqnburrXII}), and $\lambda=\alpha\mu^{1/\beta}$. Given
$\lambda
$, $\alpha$ is a power function of $\mu$. With field data only, we can
only estimate $\lambda$, which results in identifiability issues for
$\alpha$ and $\mu$. This happens in bio and medical statistics [\citet
{Hanagal11}]. In our problem, however, $\alpha$ can be estimated from
ALT data, after which $\mu$ is uniquely determined. Therefore, our
problem is free of the identifiability issue. In addition, the equality
of $\beta$ provides us a means to collate the correctness of the gamma
frailty model. Details of the procedure are as follows.

\begin{longlist}[\textit{Step} 1.]
\item[\textit{Step} 1.] Fit the lab test data using the Weibull model with c.d.f.
given by (\ref{eqnweibull}). To underscore the fact that the shape
parameter $\beta$ is estimated from the lab data, we replace it with
$\beta_L$ in the following presentation. The maximum likelihood (ML)
estimate of $(\alpha, \beta_L)$, denoted as $(\hat\alpha, \hat
\beta
_L)$, is obtained by maximizing the log-likelihood function (up to a constant)
%
%
\begin{equation}
\label{eqnwbllikelihood} l_L(\alpha, \beta_L|
\mathrm{Lab\ Data}) = \sum_{i=1}^n{ \bigl[
\delta _i(\ln\beta_L +\beta_L\ln
x_i -\beta\ln\alpha_L)-(x_i/\alpha
)^{\beta
_L} \bigr]}.
\end{equation}
Assess goodness of fit of the Weibull model. If the Weibull
distribution provides a good fit to the lab data, then proceed to step 2.
\item[\textit{Step} 2.] Fit the field return data with the Burr-XII
distribution. Here, $\beta$ in (\ref{eqnburrXII}) is replaced with
$\beta_W$ to stress the fact that this parameter is estimated from
field data. The ML estimate of $(\lambda, \beta_W, k)$, denoted as
$(\hat\lambda, \hat\beta_W, \hat k)$, is obtained by maximizing the
log-likelihood function (up to a constant)
%
%
\begin{eqnarray}
\label{eqnwbllikelihood} &&l_W(\lambda, \beta_W, k|
\mathrm{Field\ Data})\nonumber\\
&&\qquad=\sum_{j=1}^N\tilde {
\delta }_j \bigl\{\ln(k\beta_W)+\beta_W
\ln(t_j/\lambda)-\ln\bigl[(t_j/\lambda
)^{\beta_W}+1\bigr] \bigr\}
\\
&&\qquad\quad{}- \sum_{j=1}^Nk\ln
\bigl[(t_j/\lambda)^{\beta_W}+1\bigr].\nonumber
\end{eqnarray}
Assess the goodness of fit of the Burr-XII distribution. If it
provides a good fit, proceed to step 3.
\item[\textit{Step} 3.] Test the hypothesis $H_0\dvtx k = 1$ versus the
alternative hypothesis $k\neq1$. If we accept the null hypothesis, the
frailty follows an exponential distribution and the field failure time
follows a log-logistic distribution, so we can fit the field data with
the log-logistic distribution. If the hypothesis is rejected, stick to
the Burr-XII distribution.
\item[\textit{Step} 4.] Test the hypothesis $H_0\dvtx \beta_L = \beta_W$ versus
the alternative hypothesis $\beta_L \neq\beta_W$. If the null
hypothesis is accepted, then there are statistical evidences that the
frailty follows a gamma/exponential distribution, and then we can
proceed to step 5.
\item[\textit{Step} 5.] Estimate the parameters in the gamma frailty model
(\ref
{eqnfieldDistrib}) by combining the lab test data and field return
data. The ensemble log-likelihood function is
%
%
\begin{equation}\label{e10}\qquad 
l(\alpha,\beta,\lambda, k|\mathrm{All\ Data}) =  l_L(\alpha, \beta
|\mathrm{Lab\ Data})+ l_W(\lambda, \beta, k|\mathrm{Field\ Data}).
\end{equation}
\end{longlist}

To test the hypothesis in step 3, we can use either the likelihood
ratio test or the score test. These two tests are expected to be
accurate, as the size of field return data is often large. However,
these two tests may not be accurate enough when testing the hypothesis
in step 4, insofar as the lab test data are often limited. When both
the ALT data and the field return data are complete or Type II
censored, the following theorem shows that $\hat\beta_L/\hat\beta
_W$ is
a ``pivotal statistic''---that is, its distribution is independent of
the unknown parameters $\alpha, \lambda$ and $\beta$. This ratio and
its distribution will therefore be helpful in testing the hypothesis
that $\beta_L=\beta_W$ in step~4.

\begin{theorem}\label{thmpivotal}
Suppose the lab failure times follow a Weibull distribution given by
(\ref{eqnweibull}), while the field failure times conform to a
Burr-XII distribution given in~(\ref{eqnburrXII}). Consider the
hypothesis $H_0\dvtx \beta_L = \beta_W \equiv\beta$ versus the
alternative hypothesis $\beta_L \neq\beta_W$ and assume the parameter
$k$ in (\ref{eqnburrXII}) is known. When both the lab test data and
field failure data are complete or Type II censored, $\hat\beta
_L/\hat
\beta_W$ is a pivotal statistic independent of $(\alpha, \lambda,
\beta)$.
\end{theorem}

Proof of this theorem is in the \hyperref[app]{Appendix}. The proof is based on the
fact that $\hat\beta_L /\beta$ and $ \hat\beta_W /\beta$ are pivotal
statistics under the Type II censored (or complete) lab data and field
data, respectively. The constant $k$ assumption is meaningful for the
log-logistic distribution where $k=1$. When $k\neq1$ and $k$ is
estimated from field data, we can treat $\hat k$ as the true value of
$k$. This approximation should work well because the field data are
often abundant and, thus, the estimation error of $k$ is small. Theorem
\ref{thmpivotal} is not restricted by the problem of limited ALT data
and, hence, it is expected to perform better for testing the hypothesis
of $\beta_L = \beta_W$ compared with the likelihood ratio test. In a
real-life application, we would recommend conducting both tests.
When\vadjust{\goodbreak}
the results of both tests tally, there is sufficient evidence to accept
or reject the hypothesis. When the results differ, we shall stick to
the test based on Theorem \ref{thmpivotal}. The distribution of $\hat
\beta_L/\hat\beta_W$ can be obtained through simulation as follows.

\begin{algo}\label{algo1}
\begin{longlist}[1.]
\item[1.] Generate $n$ samples from Weibull(1, 1) and $N$ samples from
BXII$(1,1,k)$. For Type II censoring, the number of events will be the
same as the number of events in the data sets. For Type I censoring,
the expected number of events will be the same as the number of events
in the data sets.

\item[2.] Estimate $\hat\beta_L^*$ and $\hat\beta_W^*$ from these two data
sets separately.

\item[3.] Repeat the above two steps $B$ times to get $\hat\beta
_L^{*i}/\hat
\beta_W^{*i}$, $i=1,2,\ldots,B$.

\item[4.] Use the $B$ samples to estimate the empirical c.d.f. and sample
quantiles of $\hat\beta_L/\hat\beta_W$.
\end{longlist}
\end{algo}

In Algorithm \ref{algo1}, one can use $\hat{k}$ as the value of $k$ in the
simulation. The performance of this substitution will be evaluated
through simulation. During the lab test, both Types~I and~II
censoring are common. For warranty return data, Type~I censoring or
progressive Type I censoring are more common due to staggered entries
and warranty limits. Under this scenario, $\hat\beta_L/\hat\beta_W$ is
an approximate pivotal.

\begin{theorem}\label{thmapproximate}
Suppose the lab test data follow a Weibull distribution given by~(\ref
{eqnweibull}), while the field failure data conform to a Burr-XII
distribution given in~(\ref{eqnburrXII}). Consider the hypothesis
$H_0\dvtx \beta_L = \beta_W \equiv\beta$ versus the alternative
hypothesis $\beta_L \neq\beta_W$ and assume the parameter $k$ in
(\ref
{eqnburrXII}) is known. When the lab test data and/or field failure
data are Type~I censored, then $\hat\beta_L/\hat\beta_W$ is an
approximate pivotal statistic.
\end{theorem}

Under Type I censoring, the distribution of $\hat\beta_W/\beta_W$
depends on the unknown fraction failing at the censoring time [e.g.,
\citet{jeng01}]. Thus, it is an approximate pivotal. The approximation
improves as the sample size increases. Because the sample size of field
return data is often large, the performance of the approximation of
$\hat\beta_W/\beta_W$ is typically satisfactory. On the other hand,
according to the type of lab test data, we have the following two discussions:
\begin{itemize}
\item When the lab test data is Type II censoring, then $\hat\beta
_L/\beta$ is an exact pivotal. Thus, $\hat\beta_L/\hat\beta
_W=(\hat\beta
_L/\beta)/(\hat\beta_W/\beta)$ is an approximate pivotal because
$\hat
\beta_W/\beta_W$ is an approximate pivotal.

\item When the lab test data is Type I censoring, then $\hat\beta
_L/\beta$ is an approximate pivotal. Thus, $\hat\beta_L/\hat\beta
_W=(\hat\beta_L/\beta)/(\hat\beta_W/\beta)$ is also an
approximate pivotal.
\end{itemize}
Algorithm \ref{algo1} can still be used to do the test and the performance will
be evaluated by simulations in the next subsection.\vadjust{\goodbreak}

\subsection{Simulation study}\label{secsimulation}
In this section we conduct simulation studies to show the performance
of the statistics proposed in Theorems \ref{thmpivotal} and \ref
{thmapproximate}. In particular, we consider three scenarios:
\begin{itemize}
\item Scenario I: Type II censoring for lab data and Type II censoring
for field data.

\item Scenario II: Type II censoring for lab data and Type I censoring
for field data.

\item Scenario III: Type I censoring for lab data and Type I censoring
for field data.
\end{itemize}
We assume that the ALT uses 10 testing units whose lifetime follows a
Weibull distribution. For Scenarios I and II of the simulation, the
test is run until 8 of the units fail (i.e., Type II censoring). For
Scenario III, the expected number of failures is 8 out of 10 testing
units in the ALT (the censoring time is 733 in the simulation). For the
field data, $N$ units of the same product are sold to customers and the
environmental frailty follows $\operatorname{Gamma} (k,\mu)$. For the Type II
censoring setting (Scenario I), we stop the follow-up when $0.1N$
failures have been observed. For the Type I setting (Scenarios II and
III), the failure times are censored at $\tau$. The censoring time
$\tau
$ is so chosen that the expected proportion of field failures is 10\%.
In the simulation, we use $\alpha=534,k=1,\mu=19$ and $\tau=878$. We
examine $N=2000, 5000$ and $\beta=1.5, 2.0$.

Under each combination of $(\beta,N)$, we replicate the simulation
$2000$ times. In each replication, we compute the likelihood ratio
statistic and the statistic in Theorem \ref{thmpivotal}. The
hypothesis is rejected or accepted according to the $\tilde\alpha$
level. The estimated Type I error is obtained as the proportion of
incorrect rejections. To obtain the distribution of the pivotal, we use
$B=5000$ in each run. In the simulation, we use normal approximation
to simulate $\hat\beta_W^{*i}$ and use the distribution of $\hat
\beta
_W^{*i}/\hat\beta_W$ to approximate the distribution of $\hat\beta
_W/\beta_W$. In particular, $\hat\beta_W^{*i}$ is simulated from
$\mathcal{N} (\hat\beta_W, \sigma^2_{\hat\beta_W} )$, where
$\sigma^2_{\hat\beta_W}$ is the large sample approximate variance
estimate of $\hat\beta_W$.

Table~\ref{tabsimulatecp} shows the estimated Type I errors of the
test procedure in Theorem~\ref{thmpivotal} and the likelihood ratio
test procedure, under three different scenarios. The nominal Type I
errors that are considered in the simulation are $\tilde\alpha=0.1,
0.05$ and $0.01$. Under all scenarios, the estimated Type I errors of
the testing procedure in Theorem \ref{thmpivotal} are closer to the
nominal ones compared with the likelihood ratio statistic. In addition,
the magnitude of $N$ tends to have little effect on the Type I errors
of the likelihood ratio statistic. This is best explained by our
conjecture that the bias/error of the likelihood ratio statistic is
attributed to the small lab testing samples. Overall, we can see that
the performance of the approximate pivotal is satisfactory.

\begin{table}
\def\arraystretch{0.9}
\caption{Estimated Type I error of the test procedure in
Theorem~\protect\ref{thmpivotal} and the likelihood ratio test
procedure, under three different scenarios. The nominal Type I errors
are $\tilde\alpha=0.1, 0.05, 0.01$}
\label{tabsimulatecp}
\begin{tabular*}{\textwidth}{@{\extracolsep{\fill}}lcccccccc@{}}
\hline
&&& \multicolumn{2}{c}{$\bolds{\tilde\alpha=0.1}$} & \multicolumn{
2}{c}{$\bolds{\tilde\alpha=0.05}$} & \multicolumn{2}{c@{}}{$\bolds{\tilde\alpha
=0.01}$} \\[-6pt]
&&& \multicolumn{2}{c}{\hrulefill} & \multicolumn{2}{c}{\hrulefill} & \multicolumn{2}{c@{}}{\hrulefill} \\
{\textbf{Scenarios}}&{$\bolds{\beta}$}&{$\textbf{\textit{N}}$} & \textbf{Thm} & \textbf{LR} & \textbf{Thm} & \textbf{LR} & \textbf{Thm} & \textbf{LR} \\
\hline
{Scenario I}
&1.5& 2000 & 0.108 & 0.142 & 0.057 & 0.072 & 0.010 & 0.016 \\
&1.5& 5000 & 0.100 & 0.149 & 0.058 & 0.078 & 0.008 & 0.024 \\[3pt]
&2.0& 2000 & 0.098 & 0.134 & 0.047 & 0.078 & 0.012 & 0.014 \\
&2.0& 5000 & 0.101 & 0.138 & 0.052 & 0.071 & 0.012 & 0.020 \\[6pt]
{Scenario II}
&1.5& 2000 & 0.108 & 0.128 & 0.056 & 0.077 & 0.014 & 0.016 \\
&1.5& 5000 & 0.092 & 0.136 & 0.048 & 0.080 & 0.010 & 0.018 \\ [3pt]
&2.0& 2000 & 0.110 & 0.135 & 0.056 & 0.075 & 0.014 & 0.018 \\
&2.0& 5000 & 0.110 & 0.146 & 0.058 & 0.082 & 0.014 & 0.022 \\[6pt]
{Scenario III}
&1.5& 2000 & 0.101 & 0.117 & 0.044 & 0.062 & 0.004 & 0.012 \\
&1.5& 5000 & 0.099 & 0.126 & 0.050 & 0.062 & 0.004 & 0.014 \\[3pt]
&2.0& 2000 & 0.100 & 0.116 & 0.048 & 0.060 & 0.008 & 0.012 \\
&2.0& 5000 & 0.094 & 0.118 & 0.040 & 0.060 & 0.008 & 0.012 \\
\hline
\end{tabular*}
\end{table}
%

\section{Optimal accelerated life tests}\label{secoptimalALT}
Over the course of product evaluation and customer feedback, the
manufacturer will generate a number of design changes and come up with
a new vintage. ALTs can again be used to evaluate reliability of this
new vintage by making use of the frailty information obtained from
joint analysis of lab data and field data of previous generations. The
ALTs need to be conducted within stringent cost and time constraints,
and the testing samples need to be used efficiently. In addition, the
heterogeneous field conditions should be taken into account when
estimating life characteristics of interest. It is expected that the
operating conditions and the effects of the environments on the new
generation be approximately the same. This implies that the new vintage
will have the same frailty $Z$ with the old generation. Based on this
fact, optimal ALT plans can be developed.

For the new generation of interest, suppose its lifetime $X$ under the
stable lab testing conditions follows a Weibull distribution specified
by (\ref{eqnweibull}). Let $S_0$ be the \textit{nominal} design stress
(say, the same as the old generation) and $S_H$ be the highest
allowable test stress that has been pre-specified. For convenience, the
stress is re-parameterized as
$
\xi=(S-S_H)/(S_0-S_H).
$
It is noted that under the nominal stress $S_0$, $\xi=1$. When $X$
follows a Weibull distribution, $Y=\ln X$ conforms to a smallest
extreme value distribution with the location parameter $\eta=\ln
\alpha
$ and the scale parameter $\sigma=1/\beta$. Following the convention of
ALT design for the Weibull distribution [e.g., \citet{416}, Chapter~17],
we work with the extreme value distribution whose Fisher
information\vadjust{\goodbreak}
matrix has a closed form, and assume that the scale parameter $\sigma$
is a constant independent of the stress while the location parameter
$\eta$ depends on the stress through a linear stress-life model
\[
\eta(\xi)=\upsilon_0+\upsilon_1 \xi.
\]

Usually, the optimal test plans use only two test stresses with the
higher stress being the highest allowable stress $S_H$. Therefore, an
ALT plan is specified by the lower stress level $\xi_L$ and the
proportion of units $\pi$ for this stress. The combination $(\xi_L,
\pi
)$ is called a test plan. The purpose of the ALT design is to find out
the optimal test plan $(\xi_L^*, \pi^*)$ in order to optimize a certain
index of interest. When we are interested in life characteristics under
the \textit{nominal} conditions (i.e., characteristics based on $X$) and
ignore the heterogeneous field operating conditions, optimal
constant-stress ALTs for the extreme value distribution have been well
studied, for example, see \citet{957} for the optimal Type I censoring
plan and \citet{escobar86} for the Type II censoring case. In the
presence of the heterogeneities, however, the criteria of ALT planning
should be based on field failure times $T$ and, thus, the existing
plans are no longer optimal. Optimal plans that take the frailty $Z$
into account will be developed in this section. Denote $\bI(\xi_L,
\pi
)$ as the Fisher information matrix for $(\upsilon_0, \upsilon_1,
\sigma
)$ under the test plan $(\xi_L, \pi)$. The matrix $\bI(\xi_L, \pi
)$ has
been derived by \citet{957} under the Type I censoring scheme and by
\citet{escobar86} with Type II censoring. Therefore, use will be
directly made of these existing results.

\subsection{Minimization of the asymptotic variance of the
$p$-quantile}\label{secALTquantile}
Consider the common criterion of ALT planning that minimizes the
asymptotic variance of the ML estimator $\hat t_p$ of the $p$ quantile
of field failure time $T$. Based on (\ref{eqnfieldDistrib}) with
$\gamma=0$, the $p$-quantile of $T$ is given by
%
%
\begin{equation}
\label{eqnquantile} t_p=\alpha \bigl[ \mu(1-p)^{-1/k} - \mu
\bigr]^{1/\beta} = \exp (\upsilon_0+\upsilon_1)
\bigl[\mu(1-p)^{-1/k} - \mu \bigr]^\sigma.
\end{equation}
The asymptotic variance of the ML estimator $\hat t_p$ is $\mathit{AV}(\hat t_p)
= (\nabla t_p)^\prime\bI(\xi_L, \pi)\nabla t_p$, where $\nabla t_p$ is
the first derivative of $t_p$ with respect to $(\upsilon_0, \upsilon_1,
\sigma)$. The expression of $\nabla t_p$ is quite involved.
Alternatively, it is not difficult to show that minimization of
$\mathit{AV}(\hat t_p)$ amounts to minimization of the asymptotic variance of
$\ln\hat t_p$, which is equivalent to minimizing the asymptotic
variance of $\hat t_p/t_p$. The asymptotic variance of $\ln\hat t_p$ is
$\mathit{AV}(\ln\hat t_p) = (\nabla\ln t_p)^\prime\bI(\xi_L, \pi)\nabla
\ln
t_p$, where $\nabla\ln t_p$ is the gradient of $\ln t_p$ with respect
to $(\upsilon_0, \upsilon_1, \sigma)$ as
%
%
\begin{eqnarray}
\nabla\ln t_p(1) &=& \frac{\partial\ln t_p}{\partial\upsilon_0} = 1,
\nonumber
\\
\nabla\ln t_p(2) &=& \frac{\partial\ln t_p}{\partial\upsilon_1} = 1,
\\
\nabla\ln t_p(3) &=& \frac{\partial\ln t_p}{\partial\sigma} = \ln \bigl[
\mu(1-p)^{-1/k} - \mu \bigr].
\nonumber
\end{eqnarray}
Optimal test plans can be obtained by minimizing $\mathit{AV}(\ln\hat t_p)$
under some constraints, for example, time constraint, budget constraint
or sample size constraint.

\subsection{Minimization of the asymptotic variance of the failure
probability}
The $p$-quantile criterion considered above is often used to determine
a suitable warranty period for a new product [\citet{835}]. For a
product with a given warranty period $\tau$, what the manufacturer is
most concerned with is the proportion of field returns within $\tau$.
Therefore, another rational planning criterion is to minimize the
asymptotic variance of $\hat p_\tau$,
the ML estimate of the probability of warranty failures. This
probability is given by
%
%
\begin{equation}
\label{eqnALTprob} p_\tau=1- \bigl[ (\tau/\alpha)^\beta/\mu+ 1
\bigr]^{-k} = 1 - \biggl[ \frac{[\tau\exp(-\upsilon_0-\upsilon_1)]^{1/\sigma}}{\mu} + 1 \biggr]^{-k}.
\end{equation}
The first derivative of $p$ with respect to $(\upsilon_0, \upsilon_1,
\sigma)$ can be obtained as
%
%
\begin{eqnarray}
\nabla p_\tau(1)=\frac{\partial p}{\partial\upsilon_0} &=& -\frac
{k\Omega^{1/\sigma}}{\mu\sigma} \biggl(
\frac{\Omega^{1/\sigma
}}{\mu} + 1 \biggr)^{-k-1},
\nonumber
\\
\nabla p_\tau(2)=\frac{\partial p}{\partial\upsilon_1} &=& -\frac
{k\Omega^{1/\sigma}}{\mu\sigma} \biggl(
\frac{\Omega^{1/\sigma
}}{\mu} + 1 \biggr)^{-k-1},
\\
\nabla p_\tau(3)=\frac{\partial p}{\partial\sigma} &=& -\frac
{k\Omega
^{1/\sigma}}{\mu\sigma^2} \biggl(
\frac{\Omega^{1/\sigma}}{\mu} + 1 \biggr)^{-k-1}\ln\Omega,
\nonumber
\end{eqnarray}
where $\Omega=\tau\exp[-(\upsilon_0+\upsilon_1)]$. Based on the delta
method, the asymptotic variance is $\mathit{AV}(\hat p_\tau) = (\nabla p_\tau
)^\prime\bI(\xi_L, \pi)\nabla p_\tau$. Optimal test plans can be
determined by minimizing this asymptotic variance subject to possible
constraints on available resources.

\begin{table}[b]
\caption{Ordered failure time data observed from the ALT test}\label{tabALTdata}
\begin{tabular*}{\textwidth}{@{\extracolsep{\fill}}lccccccc@{}}
\hline
99 & 141 & 163 & 300 & 350 & 523 & 602 & 687 \\
\hline
\end{tabular*}
\end{table}

\section{Illustrative example}\label{secexample}
\subsection{Weibull fit to lab test data}
10 units of Appliance B were subject to a lab test. The experiment
ended at $t=687$ units of time, upon which 8 failures were observed and
2 were censored. In order to demonstrate Theorem \ref{thmpivotal}, we assume that the
experiment ended when the 8th failure is observed. After this
modification, the data are Type II censored. The observed failure times
of the 8 failed samples are presented in Table~\ref{tabALTdata}.

\begin{figure}

\includegraphics{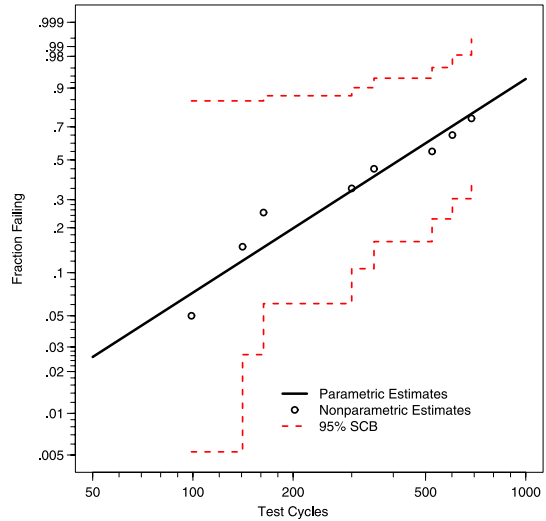}

\caption{Weibull probability plot showing the Weibull fit to the lab
data and the 95\% nonparametric SCB.}\label{figweibullfit}\vspace*{-3pt}
\end{figure}

We use the Weibull model to fit the ALT data, and the ML estimates
(standard errors) of the two parameters are $\hat\alpha=529.4$ (121.0)
and $\hat\beta_L=1.55$ (0.470), respectively. In order to visualize the
goodness of fit, we also fit the data using the Kaplan--Meier method.
The estimated c.d.f.s by means of the Weibull model and the Kaplan--Meier
method are depicted in Figure~\ref{figweibullfit}. As can be seen from
this figure, the~estimated Weibull c.d.f. passes through the empirical c.d.f.
and falls well within the 95\% simultaneous confidence band (SCB).
Therefore, the Weibull model is considered as an appropriate model for
the product under nominal conditions.\vspace*{-3pt}

\begin{figure}

\includegraphics{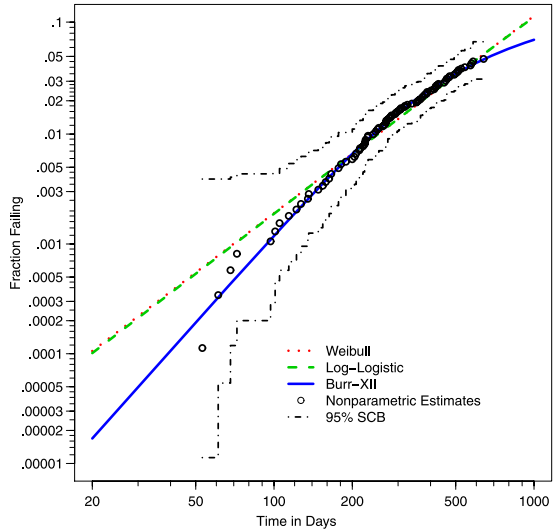}

\caption{Weibull probability plot showing the ML estimates of the
Weibull, log-logistic and Burr-XII fits to the field data and the 95\%
nonparametric SCB.}\label{figwarrantyfit}
\end{figure}

\subsection{Burr-XII fit to the field failure data}
We first use the Weibull distribution to fit the field return data. The
maximum log-likelihood value is $-977.2$. The estimated Weibull c.d.f. as
well as the empirical c.d.f. using the Kaplan--Meier method is shown in
Figure~\ref{figwarrantyfit}. As can be seen from this figure, the
Weibull distribution cannot capture the curvature of the nonparametric
estimates in the lower tail. We suspect that the inconsistency between
failures in the lab and in the field is caused by heterogeneous
operating conditions. Therefore, the gamma frailty model is invoked to
solve the problem.

We apply (\ref{eqnfieldDistrib}) to fit the data and use the
likelihood ratio statistic to test the threshold parameter $\gamma=0$.
The test reveals no evidence to reject the hypothesis. Therefore, we
set $\gamma=0$ in the following analysis. We apply the Burr-XII
distribution to fit the data, and the maximum log-likelihood value is
$-973.8$. The estimated values of the parameters are $\hat\lambda
=298.6$ (83.9), $\hat\beta_W=2.66$ (0.452) and $\hat k=0.0223$\vadjust{\goodbreak}
(0.0109), respectively. We use the Akaike information criterion (AIC)
to compare the Burr-XII model and the Weibull model for the field data.
The AIC is specified by $\mathrm{AIC}=-2l+2m$, where $l$ is the maximum
log-likelihood value of a model and $m$ is the number of parameters in
the model. The respective AIC values for the Weibull and the Burr-XII
distributions are 1958.4 and 1953.5. The Burr-XII distribution has a
smaller AIC value, indicating a better fit. As can be seen from
Figure~\ref{figwarrantyfit}, the Burr-XII distribution captures the
curvature of the nonparametric estimates in the lower tail very well,
indicating a better fit than the Weibull distribution.

We also fit the data by using the log-logistic distribution, leading to
a maximum likelihood value of $-977.0$. This value is very similar to
that of the Weibull model. Overall, the analysis suggests that the
Burr-XII distribution is more appropriate for the field data than the
Weibull model.

\subsection{The gamma frailty model}
As can be seen from the above analysis, $\hat\beta_L$ is quite close to
$\hat\beta_W$. We apply the statistic developed in Theorem \ref
{thmpivotal} to quantitatively check the correctness of the gamma
frailty model by testing $H_0\dvtx \beta_L = \beta_W$. It is easy to see
that $\hat\beta_L /\hat\beta_W = 0.585$. By making use of Algorithm \ref{algo1},
the $p$-value is 0.217. We then apply the likelihood ratio test. The
likelihood ratio statistic is 1.356 with a $p$-value of 0.244. Both
tests suggest that there is no reason to reject this hypothesis.
Therefore, we can believe that the discrepancies between the lab test
data and the field data can be explained by the frailty model, and the
gamma frailty model is appropriate for the problem.

\begin{figure}

\includegraphics{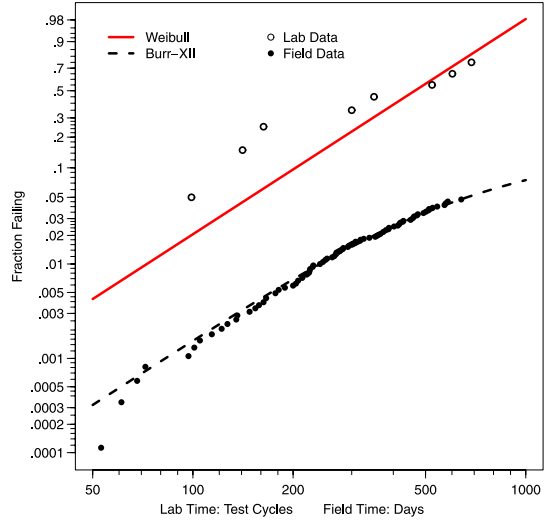}

\caption{Weibull probability plot showing the ML estimates of Weibull
fit to lab data and Burr-XII fit to the field data with a common $\beta
$.}\label{figrefit}
\end{figure}

At the last step, we estimate the parameters in (\ref
{eqnfieldDistrib}) by combining both the ALT data and the field
failure data. The ensemble of the likelihood function consists of the
Weibull likelihood contributed from the lab data and the Burr-XII
likelihood contributed from the field data. Maximization of this
function yields the ML estimates of the four parameters (standard
errors) as
$\hat\alpha=545.15$ (84.7), $\hat\beta=2.28$ (0.32), $\hat\lambda
=385.05$ (136.5) and $\hat k=0.0341$ (0.019).
Using the invariance property of the MLE, the estimated scale parameter
for the gamma frailty is $\hat\mu= (\hat\lambda/\hat\alpha)^{\hat
\beta
} = 0.452$ with a standard error 0.23. The estimated c.d.f.s for the lab
failure time distribution and the field failure time distribution can
be updated based on these parameter estimates, as shown in Figure~\ref{figrefit}.

\subsection{Optimal ALT plans}
\begin{figure}

\includegraphics{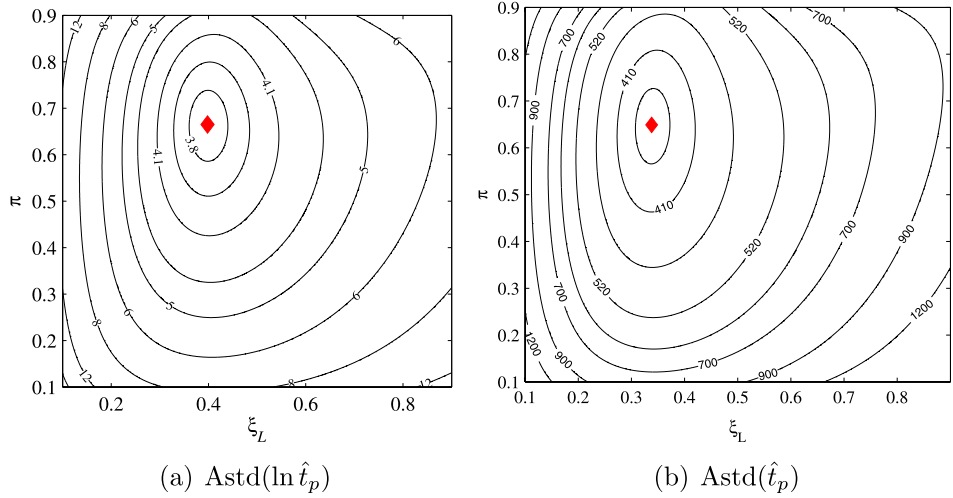}

\caption{Contours of the asymptotic standard deviation for the
two-stress optimum ALT plan.}\label{figoptimalALT}
\end{figure}

In order to improve product reliability and cater to market changes,
the manufacturer may make a number of changes to the product and come
up with a new generation. The new generation, if sold to the market,
would be operated under the same environments as the old ones and the
environments will have the same effect on the product failures.
Therefore, we assume the frailty $Z$ follows the same gamma
distribution $\operatorname{Gamma}(k,\mu)$ with $\mu=0.452$ and $k=0.0341$. Suppose
that the manufacturer is interested in knowing the 5\% quantile of the
field lifetime of the new generation, and a maximum test time of 50 is
allowed for the ALT. During the test, all units are run simultaneously.
Assume that the lifetime of the new vintage follows a Weibull
distribution under the nominal use condition, and the planning values
of the ALT are $\upsilon_0=3$, $\upsilon_1=3.4$ and $\beta=2.28$. Based
on the above settings, optimal test plans can be obtained by
numerically optimizing the asymptotic variance given in Section~\ref{secALTquantile}. For example, if the objective is to minimize the
asymptotic standard deviation, that is, the square root of the
variance, of $\ln\hat t_p$, then the optimal test plan is $(\varepsilon
^*,\pi^*)=(0.338, 0.649)$ and the associated minimal standard deviation
is 3.23. Figure~\ref{figoptimalALT}(a) shows the contour of the
asymptotic standard deviation with respect to $\varepsilon$ and $\pi$.
This test plan also minimizes the asymptotic standard deviation of
$\hat t_p$, as can be seen from Figure~\ref{figoptimalALT}(b). If we
ignore the heterogeneous field environments, the optimal test plan will
be $(\varepsilon,\pi) = (0.419, 0.766)$, which is quite different from
$(\varepsilon^*,\pi^*)$.

%
%
%

\section{Conclusions}\label{secconclusion}
This study has explained the discrepancies between in-lab failures and
field failures through the frailty model. The frailty term of each
field unit represents the unobserved operating conditions and their
complicated effects on the product failures. In the presence of
heterogeneous operating conditions, we showed that the field failure
rate can exhibit a variety of shapes, and some units may fail very
early due to severe working conditions rather than defects. ALTs should
take these heterogeneities into account. Previous research assumed
homogeneous operating conditions, which will inevitably underestimate
the variation of the field failures and, in turn, underestimate the
proportion of field returns. In addition, test plans derived under the
homogeneity assumption may be quite different from the true optimum due
to ignorance of the heterogeneity. We overcame these deficiencies and
derived the optimal plans by considering the frailty. A procedure was
developed to obtain the frailty information and to collate the validity
of the gamma distribution for the frailty. Instead of using the
likelihood ratio statistic to test the equality of $\beta_L$ and
$\beta
_W$, we suggested the use of the statistic $\hat\beta_L/\hat\beta_W$.
This statistic is pivotal under complete or Type II censored data.
Under Type I censoring, this statistic is approximately a pivotal
quantity and its good performance is demonstrated through a simulation
study. In the supplement, we further developed the inverse Gaussian
frailty models and the uniform frailty models. These two models yield
tractable field failure distributions and supplement the class of
frailty models for linking lab failures and field failures. We also
proposed an ensemble inference procedure in consideration of all the
gamma, inverse Gaussian and uniform frailty models in the supplement
[\citet{ye13supplement}].

\begin{appendix}\label{app}
\section*{Appendix}
\subsection*{Shapes of the hazard rate function of the gamma frailty model}
Taking the first derivative of (\ref{eqnfieldFailureRate}) with
respect to $t$ yields
%
%
\begin{eqnarray}
\label{eqnFRderivative}\quad h^\prime(t) &=& \biggl[(2\beta\gamma\mu-2\gamma
\mu-k)t^\beta+ \mu \alpha ^\beta(\mu\gamma+k) (\beta-1) +
\frac{(\beta-1)\gamma t^{2\beta
}}{\alpha
^\beta} \biggr]
\nonumber
\\[-8pt]
\\[-8pt]
\nonumber
&&{}\times
\frac{\beta t^{\beta-2}}{(t^\beta+ \mu\alpha
^\beta)^2}.
\end{eqnarray}
The second term on the right-hand side of (\ref{eqnFRderivative}) is
always larger than 0. So we can focus on the first term
%
%
\begin{equation}
\qquad r(x)=(\beta-1)\gamma\alpha^{-\beta}x^2 + (2\beta\gamma\mu -2
\gamma \mu-k)x+\mu\alpha^\beta(\mu\gamma+k) (\beta-1).
\end{equation}

\textit{Case} 1. When $\beta<1$, it is easy to see that $r(x)<0$ and,
hence, $h^\prime(t)<0$.

When $\beta>1$, the minimum of $r(x)$ is achieved at the point
\[
\biggl( \frac{-2\gamma\mu(\beta-1)+k}{2(\beta-1)\gamma\alpha
^{-\beta
}},\frac{4\gamma\mu\beta(\beta-1)-k}{4(\beta-1)\gamma\alpha
^{-\beta} k^{-1}} \biggr).
\]
\textit{Case} 2. When $\gamma>0, \beta>1$ and $4\gamma\mu\beta(\beta
-1)-k<0$, we see from $\beta>1$ that
\[
4\gamma\mu\beta(\beta-1)-2k\beta<0 \qquad\mbox{so } 2\gamma\mu(\beta-1)-k<0.
\]
This means that
\[
\frac{-2\gamma\mu(\beta-1)+k}{2(\beta-1)\gamma\alpha^{-\beta}}>0 \quad\mbox{and}\quad \frac{4\gamma\mu\beta(\beta-1)-k}{4(\beta-1)\gamma\alpha
^{-\beta}
k^{-1}}<0.
\]
By noting that $r(0)=\mu\alpha^{\beta}(\mu\gamma+k)(\beta-1)>0$, we
see that when $x\geq0$, $r(x)$ is positive initially, is followed by a
negative period, and then becomes positive again. From (\ref{e10}), we see
$h^\prime(t)$ also has this positive--negative--positive sign change.
Therefore, $h(t)$ increases initially, is followed by a decreasing
period, and then increases again, that is, $h(t)$ has an N-shape.

\textit{Case} 3. When $\gamma>0, \beta>1$ and $4\gamma\mu\beta(\beta
-1)-k>0$, $r(x)$ is always greater than 0, and so is $h^\prime(t)$.
Therefore, $h(t)$ is increasing over $[0, \infty)$.

\textit{Case} 4. When $\gamma=0$ and $\beta>1$, $r(x)$ reduces to
$r(x)=-kx+(\beta-1)k\mu\alpha^\beta$. This linear function is monotone
decreasing with $r(0)>0$ and $r(\infty)<0$. Therefore, $h(t)$ is
increasing at the outset and decreasing afterwards. Hence, $h(t)$ has
an upside-down bathtub shape.

\subsection*{Proof of Theorem \protect\ref{thmpivotal}}
Before proceeding to the proof of Theorem \ref{thmpivotal}, two lemmas are first presented.
%
\begin{lemma}\label{lempivotal1}
For the Burr-XII distribution given by (\ref{eqnburrXII}), conditional
on $k$, $\hat\beta_W /\beta_W$ is a pivotal statistic under Type II or
complete data.
\end{lemma}
\begin{pf}
Let $\bx=(x_1,\ldots,x_n)$ be an ordered random sample of size
$n$ from BXII$(1,1,k)$. An ordered random sample $\bt=(t_1,\ldots,t_n)$
conforming to BXII$(\beta_W, \lambda,k)$ can be obtained by taking
$t_i=\lambda x_i^{1/\beta_W}$. Suppose the sample was censored after
the $r$th observation. The ML estimates of the parameters in the
Burr-XII distribution can be obtained by deriving the score functions,
equating them to zero, and solving for the solution. Denote the ML
estimates based on $\bx$ and $\bt$ as $(\hat\beta_{W_0}, \hat
\lambda
_0)$ and $(\hat\beta_W,\hat\lambda)$, respectively. Now, we proceed to
investigate the relationship between $(\hat\beta_{W_0}, \hat\lambda_0)$
and $(\hat\beta_W,\hat\lambda)$. The log-likelihood function based on
$\bt$, up to a constant, can be written as
\begin{eqnarray*}
L(\beta_W,\lambda)&=&r\ln\beta_W+\sum
_{i=1}^r\ln(t_i/\lambda
)^{\beta
_W}-(k+1) \sum_{i=1}^r\ln
\bigl[(t_i/\lambda)^{\beta_W}+1 \bigr]
\\
&&{}-k(n-r)\ln \bigl[(t_r/\lambda)^{\beta_W}+1 \bigr].
\end{eqnarray*}
Therefore, the ML estimator $(\hat\beta_W,\hat\lambda)$ satisfies the
following equation:
\begin{eqnarray*}
& &(k+1) \sum_{i=1}^r{\frac{(t_i/\hat\lambda)^{\hat\beta_W}\ln
(t_i/\hat
\lambda)^{\hat\beta_W}} {(t_i/\hat\lambda)^{\hat\beta_W}+1}}
-r-\sum_{i=1}^r\ln(t_i/\hat
\lambda)^{\hat\beta_W}\\
&&\qquad{}+k(n-r) \frac{(t_r/\hat
\lambda
)^{\hat\beta_W}\ln(t_r/\hat\lambda)^{\hat\beta_W}} {(t_r/\hat
\lambda
)^{\hat\beta_W}+1} =0,
\\
& &(k+1) \sum_{i=1}^r{\frac{(t_i/\hat\lambda)^{\hat\beta_W}} {
(t_i/\hat
\lambda)^{\hat\beta_W}+1}}
+ k(n-r) \frac{(t_r/\hat\lambda)^{\hat
\beta
_W}} {(t_r/\hat\lambda)^{\hat\beta_W}+1}-r=0.
\end{eqnarray*}
If we substitute $t_i=\lambda x_i^{1/\beta_W}$ into the above two
equations, we can obtain
%
%
\begin{eqnarray}
 (k+1)\sum_{i=1}^r\frac{\Lambda_i\ln\Lambda_i}{\Lambda_i+1}-
r - \sum_{i=1}^r\ln\Lambda_i
+k(n-r)\frac{\Lambda_r\ln\Lambda_r
}{\Lambda
_r+1} &=& 0,
\\
(k+1)\sum_{i=1}^r
\frac{\Lambda_i}{\Lambda_i+1}+k(n-r) \frac
{\Lambda
_r}{\Lambda_r+1} -r &=& 0,
\end{eqnarray}
where $\Lambda_i= (x_i/(\hat\lambda/\lambda)^{\beta_W}
)^{\hat
\beta_W /\beta_W }$. The left-hand sides of the above two equations are
the score functions based on the sample $\bx$. Therefore, it is readily
seen that $\hat\lambda_0=(\hat\lambda/\lambda)^{\beta_W}$ and
$\hat\beta
_{W_0}= \hat\beta_W/\beta_W$. This means that the distribution of
$\hat
\beta_W/\beta_W$ is the same as $\hat\beta_{W_0}$, which does not
depend on $\beta_W$ and $\lambda$.
\end{pf}

%
\begin{lemma}\label{lempivotal2}
For the Weibull distribution specified by (\ref{eqnweibull}), $\hat
\beta_L/\beta_L$ is a pivotal statistic under Type II or complete data.
\end{lemma}
\begin{pf}
See \citet{958}.
\end{pf}

\begin{pf*}{Proof of Theorem \ref{thmpivotal}} By using Lemmas \ref{lempivotal1} and \ref
{lempivotal2} above, we can see that $\frac{\hat\beta_W/\beta
_W}{\hat
\beta_L/\beta_L}$ is a pivotal statistic. Under the null hypothesis,
this pivotal statistic is exactly $\hat\beta_L/\hat\beta_W$, which
completes the proof.
\end{pf*}
\end{appendix}

\section*{Acknowledgments}
We would like to thank the Editor, the Associate Editor and two
referees for their helpful comments on this paper.

\begin{supplement}[id=suppA]
\stitle{Supplement to ``How do heterogeneities in operating environments
affect field failure predictions and test planning?''}
\slink[doi]{10.1214/13-AOAS666SUPP} 
\sdatatype{.pdf}
\sfilename{aoas666\_supp.pdf}
\sdescription{This supplement
develops two additional frailty models, that is, the inverse Gaussian
and the uniform frailty models. An ensemble inference procedure in
consideration of all the gamma, inverse Gaussian and uniform frailty
models is also provided.}
\end{supplement}

%
%

%


\printaddresses

\end{document}